\documentclass[aoas,MSNbibl,nameyear,secthm,seceqn,dvips]{arximspdf}
\usepackage{multirow}
\usepackage{graphicx}

\doi{10.1214/11-AOAS474}
\volume{5}
\issue{3}
\pubyear{2011}
\firstpage{1816}
\lastpage{1838}

\makeatletter

\def\bsuffix #1{#1}

\let\sv@tabnotetext\tabnotetext
  \let\sv@tabnotemark@fmt\tabnotemark@fmt
   \long\def\legend#1{{\let\tabnote@indent\leavevmode\sv@tabnotetext[]{}{#1}}}

\def\@bmisc[#1]{%
  \get@battribute{unstr}%
  \common@pub@types%
  \let\bauthor\bbl@bauthor%
  \let\bhowpublished\@firstofone%
  \def\borganization##1{{\bauthor@style ##1}}%
}

\newcommand{\eqref}[1]{(\ref{#1})}

\newproclaim{defn}{Definition}[section]
\newproclaim{ex}[defn]{Example}
\makeatother

\begin{document}
\begin{frontmatter}

\title{Estimating within-household contact networks from egocentric data\thanksref{TITL1}}
\runtitle{Estimating within-household contact networks}
\thankstext{TITL1}{Supported by the NIH/NIGMS MIDAS Grant U01-GM070749.}

\begin{aug}
\author[A]{\fnms{Gail E.} \snm{Potter}\corref{}\ead[label=e1]{gail@stat.washington.edu}},
\author[B]{\fnms{Mark S.} \snm{Handcock}\ead[label=e2]{handcock@ucla.edu}},
\author[C]{\fnms{Ira M.} \snm{Longini, Jr.}\ead[label=e3]{ira@scharp.org}}\\
and
\author[C]{\fnms{M.~Elizabeth} \snm{Halloran}\ead[label=e4]{betz@u.washington.edu}
}

\runauthor{G. E. Potter et al.}

\affiliation{University of Washington and Fred Hutchinson Cancer
Research Center, University of California, Los Angeles, University of Washington and Fred~Hutchinson Cancer
Research Center and University of Washington~and~Fred~Hutchinson Cancer
Research Center}

\address[A]{G. E. Potter\\
Statistics Department\\
University of Washington \\
Box 354322\\
Seattle, Washington 98195-4322\\
USA\\
\printead{e1}}
\address[B]{M. S. Handcock\\
Department of Statistics\\
UCLA\\
8125 Math Sciences Bldg.\\
Box 951554\\
Los Angeles, California 90095-1554\\
USA\\
\printead{e2}}

\address[C]{I. M. Longini\\
M. E. Halloran\\
Fred Hutchinson Cancer Research Center\\
1100 Fairview Ave N, M2-C200\\
Seattle, Washington 98109-1024\\USA\\
\printead{e3}\\
\hphantom{E-mail: }\printead*{e4}}

\end{aug}

\received{\smonth{6} \syear{2010}}
\revised{\smonth{4} \syear{2011}}

%
\begin{abstract}
Acute respiratory diseases are transmitted over networks of social
contacts. Large-scale simulation models are used to predict epidemic
dynamics and evaluate the impact of various interventions, but the
contact behavior in these models is based on simplistic and strong
assumptions which are not informed by survey data. These assumptions
are also used for estimating transmission measures such as the basic
reproductive number and secondary attack rates. Development of
methodology to infer contact networks from survey data could improve
these models and estimation methods. We contribute to this area by
developing a model of within-household social contacts and using it to
analyze the Belgian POLYMOD data set, which contains detailed diaries
of social contacts in a 24-hour period. We model dependency in contact
behavior through a latent variable indicating which household members
are at home. We estimate age-specific probabilities of being at home
and age-specific probabilities of contact conditional on two members
being at home. Our results differ from the standard random mixing
assumption. In addition, we find that the probability that all members
contact each other on a given day is fairly low: 0.49 for households
with two 0--5 year olds and two 19--35 year olds, and 0.36 for
households with two 12--18 year olds and two 36$+$ year olds. We find
higher contact rates in households with 2--3 members, helping explain
the higher influenza secondary attack rates found in households of this
size.
\end{abstract}

%

\vspace*{15pt}
\end{frontmatter}

\section{Introduction}

Acute infectious diseases such as influenza are spread over networks of
social contacts. The 2009 pandemic influenza A (H1N1) virus has spread
to 214 countries and caused over 18,000 deaths [WHO (\citeyear{autokey39})], and a
global avian influenza pandemic continues to pose a real and dangerous
threat. Large-scale simulation models are used to predict the spread of
the epidemic and evaluate intervention strategies, but these models are
based on simplistic and strong assumptions about human interactions.
[See Halloran et al. (\citeyear{Haletal08}), Germann et al. (\citeyear{GerKadLon}), Longini et al. (\citeyear{Lonetal05}), and Ferguson et al.
(\citeyear{Feretal06}).] For example, they assume random mixing within homes, schools,
workplaces, and communities, but these social network patterns are not
estimated from surveys of contact behavior. Eubank et al. (\citeyear{Eubetal04})
implement a more detailed agent-based simulation model based on
transportation data and activity surveys, but again the model is not
informed by contact surveys. As Mossong et al. (\citeyear{Mosetal08}) stated in their
analysis of the data motivating our methods, ``Researchers often rely
on a~priori contact assumptions with little or no empirical basis.''

These basic assumptions are also used in estimating key transmission
parameters. One such parameter is the basic reproductive number
($R_0$), the expected number of secondary infections generated by a
single infectious individual in a completely susceptible population
[Anderson and May (\citeyear{AndMay91})]. Estimating $R_0$ for acute infectious
diseases commonly assumes random contacts by age group. Goeyvaerts et
al. (\citeyear{Goeetal09}) and Wallinga, Teunis and Kretschmar (\citeyear{WalTeuKre06}) use contact data
to inform the age-based contact rates used to estimate $R_0$, but other
network structures are not taken into account. Davoudi et al. (\citeyear{Davetal09})
took a new and important step by incorporating the degree distribution
in their estimation of $R_0$ for influenza, where the degree is the
number of contacts each person makes. Random mixing within households
is also assumed when estimating secondary attack rates within
households---for example, in Longini et al. (\citeyear{IMLetal88}), Halloran et al.
(\citeyear{Haletal07}), and Yang, Longini and Halloran (\citeyear{YanLonHal07}). Britton and O'Neill
(\citeyear{BriONe02}) assume random mixing in their Bayesian method to estimate the
mean of the infectious period, the infection rate, and the probability
of social contact. Demeris and O'Neill (\citeyear{DemONe05}) develop a Bayesian method
which imputes the graph of contacts between individuals from final
outcome data. They assume random mixing (within group and between
group) and separate within-group and between-group infection rates.

Network structures such as clustering, transitivity, and variation in
degree are known to play a role in disease transmission [e.g., Hethcote
 and Yorke (\citeyear{HetYor84}), Miller (\citeyear{Mil08}), and Keeling and Eames (\citeyear{KeeEam05})]. However, the
impact of these structures on transmission models is still an open area
of exploration. We can improve existing influenza simulation models by
collecting survey data on social contact behavior, developing
methodology to infer the contact network from survey data, assessing
the impact of network structures on disease spread, and finally,
integrating the important structures into the simulation models.
Parameter estimation procedures can be improved by the same process.

We contribute to the second step in this process by developing a
parametric model to estimate within-household contact networks from
diaries of social contacts and analyzing the POLYMOD data from Belgium.
In the diaries respondents reported on their contacts to other
household members, but not on contacts between other members. This
network sampling design is called egocentric. Egocentric data includes
information on respondents and people contacted, as well as numbers and
characteristics of contacts, but the identities of the people contacted
are not collected. With such data, the probability distribution of the
entire network may be inferred if we assume
that the probability of contact depends only on individual-level
attributes, or if explicit assumptions regarding dependence are made.
We take the latter approach in this paper. Koehly, Goodreau and Morris
(\citeyear{KoeGooMor04}) discuss the use of conditional log-linear models to analyze
egocentric data. Handcock and Gile (\citeyear{HanGil}, \citeyear{HanGil10}) develop a conceptual
framework for inference of network parameters from sampled data under a
variety of sampling designs. As egocentric data is commonly and easily
collected from networks, our work is applicable to network inference in
other settings.

A number of dependencies may exist in the contact network. For example,
transitivity may be present: that is, if two household members contact
the same third member, they are more likely to contact each other. Our
observed egocentric data contain limited information about dependencies
in contact behavior: for example, they do not contain information about
transitivity. However, our data set includes a household age roster for
each respondent, so some information on dependency is available. We
observed that some respondents contact no household members, but those
who contact any household members are likely to contact all or most of
them. Thus, the raw data suggest that if a respondent contacts at least
one household member, then the probability of contacting other members
is increased. We hypothesize that some respondents were away from home
on the day they filled out the contact diary (which was mailed to them
in advance). We model a latent variable indicating which household
members are at home on a given day, thus building dependency into our
network. We assume that no contacts occur to household members who are
away from home on a~given day. We estimate age-specific probabilities
of being at home as well as age-specific probabilities of contact
conditional on both household members being at home. We test whether
contact behavior differs on weekdays versus weekends, during the Easter
holiday period versus the nonholiday period, and in small ($<$4) versus
large ($\ge$4) households. We prove identifiability of our model and
use simulated data to assess conditions for weak
identifiability.\looseness=1

\section{Data}

Our data comes from the POLYMOD study, a survey in eight European
countries of social contact behavior. Mossong et al. (\citeyear{Mosetal08}) perform
descriptive analyses of this data set and analyze mixing patterns by
age. We use the Belgian data, which was collected from 750 respondents
during March--May 2006. Hens et al. (\citeyear{Henetal09}) perform a detailed analysis
of the Belgian POLYMOD data using association rules and classification
trees. Participants were recruited by random digit dialing on fixed
telephone lines. Respondents were selected to represent the urban/rural
divide in Belgium and the populations of the three main regions
(Flemish, Walloon, and Brussels). Children were oversampled, as they
play an important role in infectious disease spread. By design, 10\% of
the sample falls in each of the child age groups (0--4, 5--9, 10--14,
and 15--19) and 6\% in each of the adult age groups (20--24, 25--29,
30--34, 35--39, 40--44, 45--49, 50--54, 55--59, 60--64, 65$+$). Survey
participants were assigned two randomly selected days (one weekday and
one weekend day) and were asked to record their social contacts between
5 a.m. and 5 a.m. the following morning. Each received a paper diary
and recorded sociodemographic information of self and household, and
characteristics of all contacts made during the day. A contact was
defined to be either a physical contact or a two-way conversation of at
least three words in the physical presence of another person. Age and
sex of the person contacted were recorded, but no other identifying
information on the contacted individual was collected.

Respondents did not record whether contacted individuals were household
members or not. However, they did record the ages of all household
members in the demographic section of the survey. In addition,
respondents recorded age and sex of each person contacted, recorded
frequency of contact with that person, and checked off all locations
where that person was contacted on the day of the survey (home, work,
school, transport, leisure, and/or other). We assume that contacts
which occurred ``at home,'' were reported as ``daily or almost daily,''
and whose age matches one of the reported ages of household members,
were indeed contacts to that member. For each household we observe a
partial contact network: we have information on ties between the
respondent and all other members, but not on contacts between other
members. By design, only one respondent per household was surveyed.

Participants also recorded the date of the diary. Roughly half of
respondents (381 of 750) filled out the first day of their diary during
the two-week Easter holiday period (April 3--17), during which schools
were closed. Nearly three quarters (545 of 750) filled out the second
day of their diary during this holiday period, and over half (365 of
750) filled out both days during the holiday period.

\begin{table}
\caption{Household size distribution in the POLYMOD data from Belgium}\label{table:hh size}
\begin{tabular}{@{}lccccccccc@{}}
\hline
Household size & \hphantom{0}1 & \hphantom{00}2 & \hphantom{00}3 & \hphantom{00}4 & \hphantom{0}5 & \hphantom{0}6 & 7 & 9 & 12 \\
Number of observations & 75 & 157 & 195 & 213 & 83 & 23 & 2 & 2 & \hphantom{0}1 \\
\hline
\end{tabular}
\vspace*{-4pt}
\end{table}

\begin{table}[b]
\vspace*{-4pt}
\tablewidth=298pt
\caption{Age composition of households of size four in the Belgian
POLYMOD survey}
\label{table:composition}
\begin{tabular*}{\tablewidth}{@{\extracolsep{\fill}}lccccc@{}}
\hline
\multicolumn{5}{@{}c}{\textbf{Age category}}& \multirow{2}{50pt}[-5.5pt]{\textbf{Number of respondents}}\\[-5pt]
\multicolumn{5}{@{}c}{\hrulefill}\\
\textbf{0--5} & \textbf{6--11} & \textbf{12--18} & \textbf{19--35} & \textbf{36}$\bolds{+}$ &\\
\hline
0 & 0 & 0 & 0 & 4 & \hphantom{0}1 \\
0 & 0 & 0 & 1 & 3 & \hphantom{0}1 \\
0 & 0 & 0 & 2 & 2 & 35 \\
0 & 0 & 0 & 3 & 1 & \hphantom{0}1 \\
0 & 0 & 0 & 4 & 0 & \hphantom{0}1 \\
0 & 0 & 1 & 1 & 2 & 23 \\
0 & 0 & 1 & 2 & 1 & \hphantom{0}1 \\
0 & 0 & 2 & 0 & 2 & 40 \\
0 & 0 & 3 & 0 & 1 & \hphantom{0}2 \\
0 & 1 & 0 & 0 & 3 & \hphantom{0}1 \\
0 & 1 & 0 & 1 & 2 & \hphantom{0}1 \\
0 & 1 & 1 & 1 & 1 & \hphantom{0}2 \\
0 & 1 & 2 & 0 & 1 & \hphantom{0}1 \\
0 & 1 & 1 & 0 & 2 & 17 \\
0 & 1 & 2 & 0 & 1 & \hphantom{0}1 \\
0 & 2 & 0 & 0 & 2 & 16 \\
0 & 2 & 0 & 1 & 1 & \hphantom{0}8 \\
0 & 2 & 0 & 2 & 0 & \hphantom{0}4 \\
1 & 0 & 1 & 0 & 2 & \hphantom{0}1 \\
1 & 1 & 0 & 0 & 2 & \hphantom{0}6 \\
1 & 1 & 0 & 1 & 1 & \hphantom{0}8 \\
1 & 1 & 0 & 2 & 0 & 12 \\
2 & 0 & 0 & 0 & 2 & \hphantom{0}2 \\
2 & 0 & 0 & 1 & 1 & 12 \\
2 & 0 & 0 & 2 & 0 & 16 \\
\hline
\end{tabular*}
\legend{Each row depicts a specific age composition by showing
the number of members in each age category. The rightmost column of the
table shows the number of respondents in households of that age
composition in the POLYMOD study in Belgium.}
\end{table}

Table~\ref{table:hh size} shows the household size distribution of our
data set. Most households are size 2, 3, or 4. To give the reader a
sense of the diversity in age composition in the data set, we display
the age composition distribution in Table~\ref{table:composition} for
households of size four only. We have divided survey respondents and
their household members into the following five categories: 0--5,
6--11, 12--18, 19--35, and 36$+$, because we believe these age groups are
likely to exhibit different contact behavior. For example, 0--5 year
olds are not yet in school in Belgium and require high levels of
contact with their parents, 6--11 year olds are in primary school, and
teenagers are even more independent than 6--11 year olds so may spend
less time at home, etc. Note that some age compositions are represented
by only one or two respondents in the survey. Of course, additional age
compositions exist in the data set for households with sizes other than
four, so there is a great deal of diversity. As we are modeling
household contact networks, we restrict our attention to respondents in
households with two or more members ($n=675$).

\begin{figure}

\includegraphics{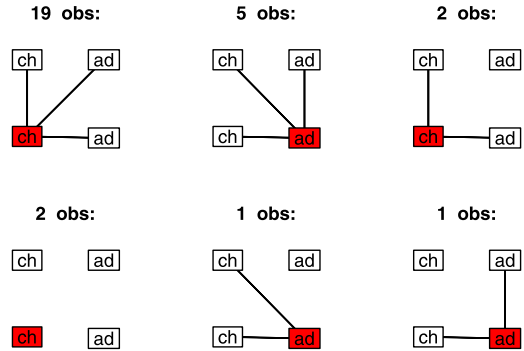}

\caption{Subset of observed data: households with two 0--5 year olds
and two 19--59 year olds; respondent in red. Lines indicate reported
contact.}
\label{fig:data}
\end{figure}

Figure~\ref{fig:data} shows a subset of the data: households of size
four with two 0--5 year olds and two 19--59 year olds. We have marked
the respondent in red. For display purposes we have assumed the two
children are exchangeable and the two adults are exchangeable. Child
respondents are likely to report making all three possible contacts,
and adult respondents are also likely to report having contacted all
three other household members. The next most likely report is two out
of three contacts. Finally, two child respondents reported contacting
no one. This seems strange as the children are 0--5 years old, but we
hypothesize that they were not at home on the day of the survey. The
paper diary mailed to respondents could be filled out anywhere, and a
parent or other guardian filled out the survey for child respondents.
We examined several types of household age compositions and always
found a subset of respondents to report no household contacts. Overall,
16\% of respondents report no household contacts, yet those who report
at least one contact contacted an average of 88\% of their household
members. This suggests a dependency in contact behavior: if at least
one household member is contacted, then others are more likely to be contacted.

\begin{figure}

\includegraphics{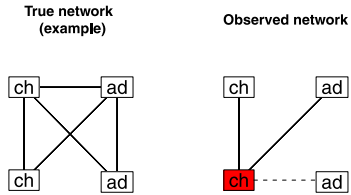}

\caption{Example of a true network and the observed portion from the
POLYMOD study. In the plot of the observed portion, the respondent is
red, solid lines indicate observed contacts, and the dashed line shows
the observed noncontact.}
\label{fig:net_example}
\vspace*{4pt}
\end{figure}

Figure~\ref{fig:net_example} shows an example of how the observed data
compare to the true, complete network. We develop a model to infer the
probability distribution of the complete network, based on partial
observations of the network.

\section{Methods}
In this section we present a model for the contact network and develop
inference for it based on the incomplete information available in the
egocentric data. The model for the contact network is of primary
scientific interest.

\subsection{A latent variable model}
\label{subsection:likelihood}

Our inspection of the observed data revealed that some respondents
reported no ``at home'' contacts to other household members on the day
of the survey. This may occur because the respondent was not actually
at home on the day of the survey, or because he/she was at home but
made no social contacts at home. Our data do not directly distinguish
between a respondent being away from home versus being at home but not
contacting any household members. We use a latent variable model to
tease apart these two phenomena.

For a household of size $k$, let $\mathbf{Z}$ denote a random matrix
representing the at home contact network. We represent $\mathbf{Z}$ by
a $k$ by $k$ sociomatrix, where
\[
Z_{ij} = \cases{
1, &\quad if there is a contact between person $i$ and person $j$,\cr
0, &\quad otherwise.
}
\]
Let $\mathbf{H}$ be a Bernoulli random vector of length $k$, indicating
whether each household member is home or not. We assume that the
elements of $\mathbf{H}$ are independent, that is, the absence of one
household member does not influence whether another household member is
also absent. If $\mathbf{H}$ is unobserved, we can express the
likelihood of $\mathbf{Z}$ by the Law of Total Probability as follows:
%
\begin{equation}
P(\mathbf{Z}=\mathbf{z})=\sum_{\mathbf{h} \in\mathcal{H}} P(\mathbf
{Z}=\mathbf{z}|\mathbf{H}=\mathbf{h})P(\mathbf{H}=\mathbf{h}).
\end{equation}
Above, $\mathcal{H}$ represents the space of all possible ``at home''
vectors $\mathbf{H}$. We now add some assumptions about the
distributions of $\mathbf{H}$ and $\mathbf{Z}|\mathbf{H}$.

We assume that $H_i \sim$ Bernoulli($p_v$), where $v$ is the age
category of household member $i$. We parametrize the distribution of
$\mathbf{Z}|\mathbf{H}$ by assuming that contacts
$Z_{ij}|(H_i=1,H_j=1)$ are independent Bernoulli random variables whose
probability parameters depend only on the age categories of household
members $i$ and~$j$. We define $p_{\mathit{rs}}$ as the probability of contact
between a~member of age category $r$ with a member of age category $s$,
conditional on both of them being at home. So $Z_{ij}|(H_i=1,H_j=1) \sim
 \operatorname{Bern}(p_{\mathit{rs}}$), where~$r$ is the age category of member $i$ and $s$
is the age category of member $j$. We assume contacts are symmetric, so
$p_{\mathit{rs}}=p_{sr}$. We will model only at-home contacts between household
members, so $Z_{ij}$ is zero when either $H_i=0$ or $H_j=0$. Thus, we
assume that the only dependence in contacts between members comes from
whether the members are at home or not.

The Bernoulli assumptions allow us to collapse contacts into counts by
age groups. Although our outcome of interest is the sociomatrix, we
observe only a single row of the sociomatrix for each household. Under
our model assumptions, a sufficient statistic for the contribution of
each household is a vector $\mathbf{W}$, with elements $W_s = $ the
number of contacts observed from the respondent to household members in
age group $s$, for $s\in\{1,\ldots,5\}$. Let $\mathbf{n} = (n_1, n_2,
n_3, n_4, n_5)$ denote the number of nonrespondent household members in
each age category. Then $n_s - W_s$ is the number of members in age
category $s$ who were not contacted by the respondent.

With a slight abuse of notation, we will still use $\mathbf{H}$ to
denote home/away status, but the elements will be counts rather than
indicators. Now let $H_{v}$ be the number of nonrespondent household
members in age category $v$ who are at home rather than the home/away
status of member $v$. The new $\mathbf{H}$ has length 5 regardless of
household size. Then $H_{v}$ follows independent binomial distributions
with parameters $n_v$ and $p_v$, where $n_v$ is the number of
nonrespondent household members in age category $v$, and $p_v$ is the
probability of a person in age category $v$ being at home. In addition,
let $R$ denote the home/away status of the respondent, with $R=1$ if
the respondent is home, and $R=0$ otherwise. Since respondents were
mailed a paper diary in advance of their survey date and returned it by
mail, and since some respondents did not list any household contacts in
their diary, the ``at home'' status of the respondent is unobserved.
The latent variables of interest are~$\mathbf{H}$ and $R$.\looseness=-1

Under these assumptions the likelihood contribution for a respondent in
age category $j$ is
%
\begin{equation} \label{lik1}
P(\mathbf{W}=\mathbf{w})=P(\mathbf{W}=\mathbf{w}|
R=0)(1-p_j)+P(\mathbf{W}=\mathbf{w}|R=1)p_j.
\end{equation}

Above, $P(\mathbf{W}=\mathbf{w}|R=0)=1$ if $\mathbf
{w}=(0,0,0,0,0)$ and zero if $w_s>0$ for at least one $s\in\{1,2,3,4,5\}
$. If the respondent is at home, it follows from our assumptions that
contacts to other household members are independent, so we can rewrite
the second term as follows:

\begin{equation} \label{lik2}
P(\mathbf{W}=\mathbf{w}|R=1)p_j = \prod_{s=1}^5 P(W_s =
w_s|R=1)p_j.
\end{equation}

Household members who had reported contact with the respondent were
necessarily at home. Those without reported contact could have been
away from home, or could have been at home but not contacted. By
applying the Law of Total Probability again, conditioning on the
home/away status of nonrespondent household members, \eqref{lik2} becomes

\begin{equation} \label{lik3}
p_j \sum_{h_1=0}^{n_1} \sum_{h_2=0}^{n_2} \cdots\sum_{h_5=0}^{n_5}
\prod_{s=1}^5
P(W_s=w_s|H_s=h_s)P(H_s=h_s).
\end{equation}

By applying our distributional assumptions, this term becomes

\begin{eqnarray} \label{lik4}
&&p_j \sum_{h_1=w_1}^{n_1} \sum_{h_2=w_2}^{n_2} \cdots\sum_{h_5=w_5}^{n_5}
\prod_{s=1}^5\pmatrix{h_s \cr
w_s}p_{js}^{w_s}(1-p_{js})^{h_s-w_s}\nonumber\\[-8pt]\\[-8pt]
&&\hphantom{p_j \sum_{h_1=w_1}^{n_1} \sum_{h_2=w_2}^{n_2} \cdots\sum_{h_5=w_5}^{n_5}\prod_{s=1}^5}
{}\times\pmatrix{n_s \cr h_s }p_s^{h_s} (1-p_s)^{n_s-h_s}.\nonumber
\end{eqnarray}

We assume that households are independent, so the likelihood of the
entire data set is the product of the likelihood contributions of all
respondents. Note that the parameters $n_s$ are determined by the data
and differ for different respondents.

To aid in understanding, we provide an example for the reader.\vspace*{4pt}

\begin{ex}
Suppose the respondent is in age group 1, and has two household
members, one in age group 2 and one in age group 4, and suppose the
respondent reports no contacts to household members on the day of the
survey. Then,
$\mathbf{W}=(0,0,0,0,0)\equiv\mathbf{0}$ and $\mathbf{n}=(0,1,0,1,0)$.
The likelihood contribution for this respondent is
\begin{eqnarray*}
P(\mathbf{W}=\mathbf{0})&=&P(\mathbf{W}=\mathbf{0}|
R=0)(1-p_1)+P(\mathbf{W}=\mathbf{0}|R=1)p_1
\\
&=&1-p_1 + p_1 \bigl( P(\mathbf{W}=\mathbf{0}|\mathbf{H}=\mathbf{0})P(\mathbf
{H}=\mathbf{0})
\\
&&\hphantom{1-p_1 + p_1 (}{}+ P\bigl(\mathbf{W}=\mathbf{0}|\mathbf{H}=(0,1,0,0,0)\bigr)P\bigl(\mathbf
{H}=(0,1,0,0,0)\bigr)
\\
&&\hphantom{1-p_1 + p_1 (}{}+ P\bigl(\mathbf{W}=\mathbf{0}|\mathbf{H}=(0,0,0,1,0)\bigr)P\bigl(\mathbf
{H}=(0,0,0,1,0)\bigr)
\\
&&\hphantom{1-p_1 + p_1 (}{}+ P\bigl(\mathbf{W}=\mathbf{0}|\mathbf{H}=(0,1,0,1,0)\bigr)P\bigl(\mathbf
{H}=(0,1,0,1,0)\bigr) \bigr)
\end{eqnarray*}
by the Law of Total Probability, the independence of $\mathbf{H}$ and
$R$, and the fact that
$P(\mathbf{W}=\mathbf{0}|R=0)=1$. Next we apply the
distributional assumptions on $P(\mathbf{W}|\mathbf{H})$ and $P(\mathbf
{H})$ to obtain
\begin{eqnarray*}
&=&1-p_1 + p_1 \bigl( (1-p_{2})(1-p_{4}) + (1-p_{12})p_2(1-p_4) \\
&&\hphantom{1-p_1 + p_1 (}
{}+
(1-p_{14})(1-p_2)p_4 + (1-p_{12})(1-p_{14})p_2 p_4 \bigr).
\end{eqnarray*}
Through algebra we can see that this is equivalent to \eqref{lik1}.
\end{ex}

\subsection{Maximum likelihood estimation}
By maximizing the likelihood we estimate the probability parameters
$p_{\mathit{rs}}$ and $p_v$ for $r,s,v \in\{1,2,3,4,5\}$. We note that a
Bayesian approach would also be appropriate for our question of
interest, as we expect contact probabilities within households to be
high, particularly when one of the members is a young child. We chose
not to use a Bayesian approach because we prefer not to increase the
subjectivity of our results.

$\!\!$Optimization was performed in R version 2.8 with the \texttt{optim}
function~[R~De\-velopment Core Team (\citeyear{Dev10})]. We used the BFGS method, a quasi-Newton method published
simultaneously by Broyden (\citeyear{Bro70}), Fletcher (\citeyear{Fle70}), Golfarb (\citeyear{Gol70}), and Shanno (\citeyear{Sha70}). The
\texttt{optim} function estimates the Hessian of the log likelihood at the
MLE, so providing an estimate of the observed Fisher information matrix
which one can invert to compute confidence intervals. However, some
parameter estimates were on the boundary of the parameter space, so we
computed confidence intervals by a nonparametric bootstrap, as
described by Efron and Tibshirani (\citeyear{EfrTib93}), instead of by inverting the
Fisher information matrix. We used 1,000 bootstrap iterations. In one
case, both lower and upper bounds of the interval were estimated to be
1 since all data points supported a parameter estimate of 1. Since the
bootstrap fails as an estimate of uncertainty in this case, we omit the
lower bound of this interval. R code used for estimation is included in
the supplementary material [Potter et al. (\citeyear{PotetalN3})]. Network graphs
were produced with statnet software [Handcock et al. (\citeyear{Hanetal03})].

\subsection{Identifiability of the latent variable model}

Since we are estimating a latent variable from a data set with
structurally missing data, it is not immediately apparent that our
parameters are identifiable. According to Silvey (\citeyear{Sil75}), a parameter is
identifiable if distinct values of the parameter vector give distinct
probability distributions on the sample space. We prove identifiability
of our parameter vector in the \hyperref[app]{Appendix}. It is possible that the
identifiability is only ``weak.'' Identifiability guarantees that the
parameter can be determined with an infinite amount of data, but ``weak
identifiability'' means that even very large data sets do not contain
enough information to precisely estimate the parameter [Bolker (\citeyear{Bol08})].
Because we are using partially observed network data to estimate 20
parameters, five of which correspond to a latent variable, it is not
immediately obvious that our data set is large and diverse enough to
disentangle the ``at home'' probabilities from the conditional contact
probabilities. We perform a simulation study to assess whether data
sets with the same size and distribution of household age compositions
as ours contain enough information to estimate our
parameters.\looseness=-1

\subsection{Model selection}

We investigated three effects which could help to model contact
behavior. First, contact probabilities may vary with household size, as
people in large households may be less likely to contact all other
members than those in small households. We also tested for differences
during the Easter holiday and a nonholiday period, and between weekend
days and weekdays. Because we are performing three statistical tests,
we applied Bonferroni's correction for multiple testing: we use a
critical value of $\alpha=0.05/3 = 0.017$ instead of $\alpha=0.05$
[Abdi (\citeyear{Abd})].

Let $p_{\mathit{rs},\mathit{small}}$ denote the conditional probability of contact
between household members in age groups $r$ and $s$ for households with
2--3 members, and~$p_{\mathit{rs},\mathit{large}}$ denote the conditional probability of
contact between household members in age groups $r$ and $s$ for
households with four or more members. Similarly, let $p_{s,\mathit{small}}$ and
$p_{s,\mathit{large}}$ be the probabilities of a member in age category $s$
being at home in small and large households. Let $\Omega_0$ be the
subspace in which we have restricted parameters for small households to
be equal to those for large households: that is,
$p_{\mathit{rs},\mathit{small}}=p_{\mathit{rs},\mathit{large}}$ and $p_{s,\mathit{small}} = p_{s,\mathit{large}}$ for $r,s\in
\{1,2,3,4,5\}$. We are interested in testing whether $\mathbf{p}\in
\Omega_0$ or $\mathbf{p}\in\Omega\setminus\Omega_0$. Because three of
the parameter estimates are on the boundary of the space ($p_{0\mbox{--}5, 0\mbox{--}5}
= 1, p_{6\mbox{--}11,6\mbox{--}11}=1$, and $p_{6\mbox{--}11,12\mbox{--}18}=1$), the conditions for the
classical likelihood ratio test using Wilk's (\citeyear{Wil38}) theorem do not
hold. However, when estimation was performed separately for small and
large households, we found $p_{0\mbox{--}5, 0\mbox{--}5} = 1, p_{6\mbox{--}11,6\mbox{--}11}=1$, and
$p_{6\mbox{--}11,12\mbox{--}18}=1$ for both small and large households. In both cases,
there is not enough variability in the data to compute a confidence
interval, suggesting that the true value is close to 1 for both small
and large households. These parameters are estimated with sample sizes
ranging from 29--34, and the data is consistent with a parameter value
of 1. For this reason we considered it unnecessary to test for a
household effect for these three parameters. Instead, we assumed that
these three parameters were equal for small and large households, and
tested whether any of the other 17 parameters differed for small versus
large households. This permits us to do a classical likelihood ratio
test, in which the test statistic is compared to a chi-square
distribution with 17 degrees of freedom. Our test statistic was 37.4
with a $p$-value of 0.003, so we concluded that one or more of the
parameters differs for small versus large households. While the
estimated ``at home'' probabilities were similar for small and large
households, nearly all conditional contact probability estimates were
larger in small households than in large households. We chose not to
include a household size effect in our final model, as some cell counts
were too small to obtain reasonable estimates. The separate estimates
for small and large households are included in the supplementary
material [Potter et al. (\citeyear{PotetalN1})].

We used the same method to assess whether contact behavior differed on
the weekend versus on a weekday. Here, only one parameter estimate was
on the boundary of the space. Our likelihood ratio test statistic was
23.3, which when compared to a chi-square distribution with 19 degrees
of freedom gives a $p$-value of 0.22. Thus, we found no evidence that
contact behavior differed over the weekend versus on a weekday.\vadjust{\eject}

Similarly, we tested the null hypothesis that the parameters were the
same during the two-week Easter holiday period as during a nonholiday
period against the alternative that one or more probability parameters
could differ between the holiday and the nonholiday. Since our test
statistic was 53.3 with a $p\mbox{-value} < 0.001$, we concluded that
within-household contact behavior in Belgium is different during the
Easter holiday period than during a nonholiday period. However, we did
not see a systematic, meaningful, and substantive pattern explaining
the difference. For this
reason, we chose not to include a holiday effect in our final model. The separate holiday and nonholiday estimates are
included in the supplementary material [Potter et al. (\citeyear{PotetalN1})].

\section{Results}

\subsection{Parameter estimates}

Table \ref{table: contact probs} shows maximum likelihood estimates for
the probability of contact between two members, conditional on them
both being at home. Table~\ref{table: home probs} shows estimates of
the probability of members being at home on a given day. We see that
contact probabilities are quite high from young children to all age
groups, and decrease slightly as the ages of both members increase.

\begin{table}
\tablewidth=300pt
\caption{Conditional contact probability estimates with 95\% bootstrap
confidence intervals}\label{table: contact probs}
\begin{tabular*}{\tablewidth}{@{\extracolsep{\fill}}lccccc@{}}
\hline
\textbf{Age}&&&&&\\
\textbf{category}& \textbf{0--5} &\textbf{6--11} &\textbf{12--18} &\textbf{19--35} &\textbf{36}$\bolds{+}$\\
\hline
0--5& 1.00 & 0.90 & 0.67 & 0.99 & 0.96 \\
& [--, 1.00]& [0.76, 0.99]& [0.24, 0.99]& [0.93, 1.00]& [0.86, 1.00]\\
&&&&&\\
6--11& & 1.00 & 1.00 & 0.96 & 0.91 \\
&& [0.86, 1.00]& [0.89, 1.00] &[0.88, 1.00] & [0.82, 0.98]\\
&&&&&\\
12--18& & & 0.88 & 0.65 & 0.91 \\
&&& [0.74, 0.99]&[0.48, 0.81] &[0.85, 0.97]\\
&&&&&\\
19--35& & & & 0.80 & 0.83 \\
&&&& [0.65, 0.94]& [0.75, 0.90] \\
&&&&&\\
36$+$& & & & & 0.89\\
&&&&& [0.81, 0.97]\\
\hline
\end{tabular*}
\end{table}

\begin{table}
\caption{Estimated probabilities of being at home with 95\% bootstrap
confidence intervals}
\label{table: home probs}
\begin{tabular}{@{}lccccc@{}}
\hline
\textbf{Age}&&&&&\\
\textbf{category}& \textbf{0--5} &\textbf{6--11} &\textbf{12--18} &\textbf{19--35} &\textbf{36}$\bolds{+}$\\
\hline
Probability & 0.90&0.92& 0.89& 0.90& 0.92\\
&[0.86, 0.95] &[0.88, 0.98]&[0.84, 0.94] &[0.86, 0.94]& [0.89, 0.95]\\
\hline
\end{tabular}
\end{table}

\begin{figure}[b]

\includegraphics{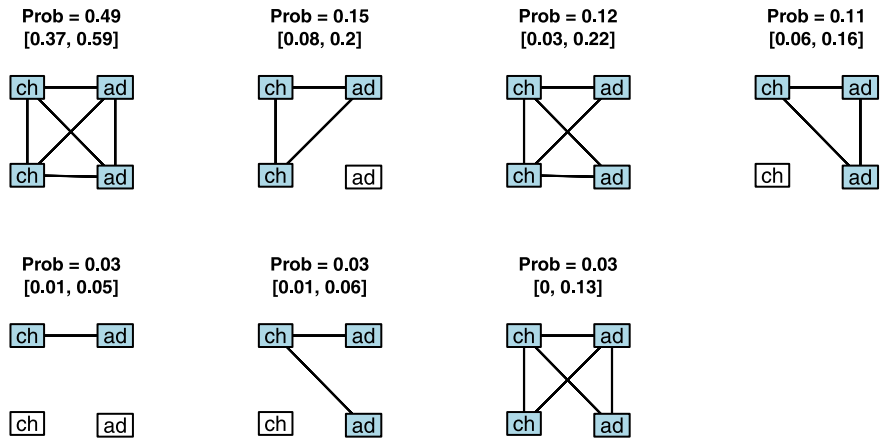}

\caption{Estimated probability distribution of contact networks,
households with two 0--5 year olds and two 19--35 year olds, with 95\%
confidence intervals. Members at home are shown in blue; those away
from home are shown in white. Only networks with probability $>0.02$
are depicted.}\label{fig:complete-nets-young-hh}
\end{figure}

Our 20 parameters and our distributional assumptions determine the
probability distribution of within-household contact networks for any
household of a specified size and age composition. Figure~\ref
{fig:complete-nets-young-hh} shows the estimated probability
distribution of contact networks for households with two 0--5 year olds
and two 19--35 year olds. The probability of the first network depicted
is the probability that all household members are at home times the
probability that all contacts between them occur. Other network
probabilities are computed similarly. Confidence intervals were
computed by performing this deterministic computation 1,000 times,
using the parameter estimates obtained from the 1,000 bootstrap
re-samples of our data set.

The ``at-home'' status of each member is indicated by color: blue
members are at home and white members are away from home. According to
our model, the most likely network includes all possible contacts,
which fits with our understanding of social behavior. This network is
estimated to have a~49\% chance of occurring on a given day in this
type of household. The second most likely network shows one of the
adults away from home, but all other contacts occurring. The third most
likely network, with probability 12\%, has all members at home, and all
contacts except the one between the two adults occurring.

\begin{figure}

\includegraphics{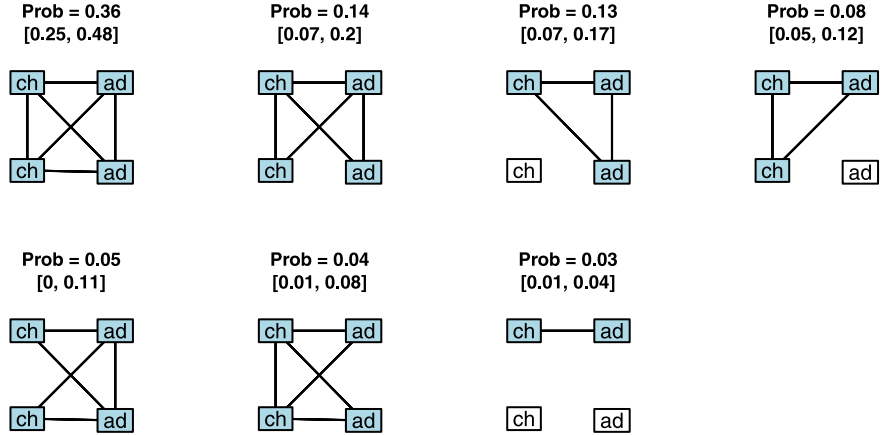}

\caption{Estimated probability distribution of contact networks,
households with two 12--18 year olds and two 36$+$ year olds, with 95\%
confidence intervals. Members at home are shown in blue; those away
from home are shown in white. Only networks with probability${}>{}$0.02
are depicted.}\label{fig:complete-nets-teen-hh}
\end{figure}

Figure~\ref{fig:complete-nets-teen-hh} shows the estimated probability
distribution for contact networks with two 12--18 and two 36$+$ year
olds. As with the younger household type, the most likely network is
the one in which all contacts occur, but its estimated probability is
0.36, rather than 0.49. As teenagers are more independent than children
under 5, this seems reasonable. The second most likely network is one
in which all members are at home, but one of the child-adult contacts
does not occur, and the third most likely network has one teenager away
from home, but all other contacts occurring. These estimates are also
reasonable given our understanding of social behavior.

\begin{figure}[b]

\includegraphics{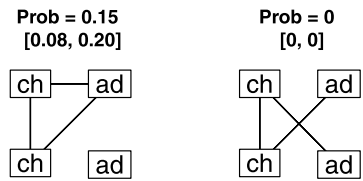}

\caption{Illustration of dependency in our model. Under random mixing
or under an independence model with age-specific contact probabilities
but no latent variable, the two networks below would have the same probability.}\label{fig:show_dependency}
\end{figure}

The dependency in our model can be seen by studying Figures~\ref
{fig:complete-nets-young-hh} and~\ref{fig:complete-nets-teen-hh}.
Networks which would be equally likely under an independence assumption
have different estimated probabilities under our assumptions. For
example, Figure~\ref{fig:show_dependency} shows two possible contact
networks and their probabilities computed under our model for a
household with two 0--5 year olds and two 19--35 year olds. Under a
random mixing assumption these two networks would have the same
probability since they have the same numbers of child--child ties and
adult--adult ties. An independence model which has age-specific contact
probabilities but no latent variable effect would also assign the same
probabilities to these two networks. Yet under our model, one of them
has probability 0.15, and the other has probability 0. Thus, our
assumptions give rise to a process very different from random mixing.
The latent variable in our model creates dependencies which would not
be captured in a model with only age-specific mixing probabilities.

\begin{table}[b]
\caption{Validity check, all households}
\label{table:validity check}
\begin{tabular}{@{}lcc@{}}
\hline
\textbf{Age}& \textbf{Estimated probability}&\textbf{\% of respondents with} \\
\textbf{category}& \textbf{of being at home}& \textbf{any at home contacts} \\
\hline
0--5 & 0.90 [0.86, 0.95] & 0.89 [0.83, 0.96] \\
6--11 & 0.92 [0.88, 0.98] & 0.93 [0.87, 0.98] \\
12--18 & 0.89 [0.84, 0.94] & 0.88 [0.81, 0.94] \\
19--35 & 0.90 [0.86, 0.94] & 0.82 [0.75, 0.89] \\
36$+$ & 0.92 [0.89, 0.95] & 0.80 [0.75, 0.85] \\
\hline
\end{tabular}
\end{table}

\subsection{Model validity and weak identifiability}

Our results suggest that our algorithm has succeeded at uncovering the
parameter values and disentangled the home/away process from the
contact process for our data set. However, it is possible that the
identifiability is only weak. In this section we show results from a
validity check evaluating our model and perform simulations to assess
weak identifiability.

To check the validity of our model, we compare our estimates of ``at
home'' probabilities to the percentage of respondents who report any
contacts to household members. Since respondents are randomly sampled,
these percentages are unbiased estimates of the probability of a person
having at least one contact to another household member at home. The
probability of being at home is greater than or equal to the
probability of contacting at least one household member at home, since
the latter event implies the former.\looseness=-1

Table \ref{table:validity check} compares MLEs of the probability of
being at home to the estimated probability of being at home and
contacting at least one household member. For 4 of 5 age groups, the
estimated probability of being at home is greater than the estimated
proportion of people contacting any household members at home, as we
expect. The difference is statistically significant only for the oldest
age group. For 6--11 year olds, the direction of the difference is
opposite of what we expect, but the statistically insignificant
difference is small enough not to raise concern. Although the
probability of being at home is necessarily greater than or equal to
the probability of contacting anyone at home, we expect these
probabilities to be close. Our validity check indicates that our model
is producing reasonable results.

\begin{table}[b]
\tablewidth=295pt
\caption{At home probabilities used for simulations, mean estimated at
home probabilities over 500 simulations, and 2.5\% and 97.5\% quantiles
of the estimates}
\label{table:sim1a}
\begin{tabular*}{\tablewidth}{@{\extracolsep{\fill}}lccc@{}}
\hline
\textbf{Age} & \textbf{Truth} & \textbf{Mean of estimates} & \textbf{95\% quantile interval} \\
\hline
0--5 & 1\hphantom{.0} & 1.00 & [1.00, 1.00] \\
6--11 & 0.9 & 0.93 & [0.87, 0.98] \\
12--18 & 0.8 & 0.85 & [0.77, 0.92] \\
19--35 & 0.7 & 0.73 & [0.68, 0.79] \\
36$+$ & 0.6 & 0.61 & [0.57, 0.64] \\
\hline
\end{tabular*}
\end{table}

We performed a simulation study to assess whether data sets with the
same size and distribution of household age compositions as ours
contain enough information to estimate our parameters. The simulation
procedure was as follows:
\begin{enumerate}
\item Choose values for the five ``at home'' probabilities and the 15
conditional contact probabilities.
\item Simulate 500 data sets with the same size and distribution of
household age compositions as ours from the model using these parameters.
\item For each simulated data set, compute maximum likelihood
estimates of the parameters.
\item Compute the mean of the MLEs over the 500 simulations and
compare to the true value.
\end{enumerate}

We performed simulations for two different sets of parameter values.
First we set the conditional contact probabilities equal to our
estimated contact probabilities, but we varied the ``at home''
probabilities in our simulation to test whether the method could detect
the variation. (Recall that all of our estimated ``at home''
probabilities were near 0.90.) We chose the values 1.0, 0.9, 0.8, 0.7,
and 0.6 for ``at home'' probabilities of the five age groups. Our
results in Tables~\ref{table:sim1a} and \ref{table:sim1b} indicate that
the estimation procedure does a good job of uncovering the true ``at
home'' probabilities, and a fair job of uncovering the conditional
contact probabilities. The accuracy of the conditional contact
probability estimates is highest when the two age groups have a high
probability of being at home. These estimates are most accurate when
one of the age groups is 0--5, whose probability of being at home is
one, and least accurate when one of the age groups is 36$+$, who have the
smallest probability (0.60) of being at home. Since our estimated ``at
home'' probabilities from the actual data are all near 0.90, our
conditional contact probability estimates are probably fairly accurate.

\begin{table}
\tablewidth=305pt
\caption{Conditional contact probabilities used for simulations, mean
estimated at home probabilities over 500 simulations, and 2.5\% and
97.5\% quantiles of the estimates}
\label{table:sim1b}
\begin{tabular*}{\tablewidth}{@{\extracolsep{\fill}}lcccc@{}}
\hline
\textbf{Age 1} & \textbf{Age 2} & \textbf{Truth} & \textbf{Mean of estimates} & \textbf{95\% quantile interval} \\
\hline
0--5 & 0--5 & 1.00 & 1.00 & [1.00, 1.00] \\
& 6--11 & 0.90 & 0.89 & [0.78, 1.00] \\
& 12--18 & 0.67 & 0.67 & [0.32, 0.99] \\
& 19--35 & 0.99 & 0.96 & [0.86, 1.00] \\
& 36$+$ & 0.96 & 0.93 & [0.81, 1.00] \\[3pt]
6--11 &6--11 & 1.00 & 0.94 & [0.79, 1.00] \\
& 12--18 & 1.00 & 0.93 & [0.78, 1.00] \\
& 19--35 & 0.96 & 0.89 & [0.76, 1.00] \\
& 36$+$ & 0.91 & 0.83 & [0.72, 0.94] \\[3pt]
12--18& 12--18 & 0.88 & 0.80 & [0.63, 0.98] \\
& 19--35 & 0.65 & 0.59 & [0.41, 0.78] \\
& 36$+$ & 0.91 & 0.80 & [0.71, 0.87] \\[3pt]
19--35 & 19--35 & 0.80 & 0.75 & [0.56, 0.99] \\
& 36$+$ & 0.83 & 0.74 & [0.64, 0.82] \\[3pt]
36$+$ & 36$+$ & 0.89 & 1.00 & [0.99, 1.00] \\
\hline
\end{tabular*}
\end{table}

Our data set contains fairly high reported rates of contact. A data set
with lower contact rates may not provide enough information to
distinguish household members being away from home versus not being
contacted. To investigate this, we performed a second simulation for
which we reduced contact probabilities to obtain empirical data sets
with households in which some respondents are home but don't contact
any other members. Our results, given in the supplementary material
[Potter et al. (\citeyear{PotetalN2})], show that in this type of data set the
procedure does not work as well. The ``at home'' probabilities are
underestimated, and the contact probabilities are overestimated.

\section{Discussion}

In this paper we infer the structure of within-household contact
networks, which are a key component for models of epidemic spread. We
show how to infer the probability distribution of the complete
within-household contact network from individual-level data from one
respondent per household in a random sample of households. By modeling
the unobserved event that some members may be away from home on a given
day, we incorporate dependency in contact behavior, resulting in a
process different from random mixing. We also find the probability of
all household members contacting each other on a given day to be
substantially less than one. These two findings indicate that contact
behavior reported in surveys is different from the contact patterns
generally used for epidemic models and estimation methods. Our finding
that contact probabilities are higher in households with 2--3 members
than in households with 4$+$ members helps to explain the higher
transmission rates found by Cauchemez et al. (\citeyear{Cauetal09}) in households with
2--3 members than in larger households.

The contact probability matrices show that contact between any two
members is highly likely if both members are at home. All probabilities
are over 50\%, and most range from 90--100\%. In any size household,
0--5 year olds are highly likely to contact other young children and
adults, as we might expect. The contact probability is lowest between
teenagers in any size household, as we might also expect. Our model
succeeded at disentangling the contact process from the home/away
process, and the estimated probabilities of being at home are all close
to 90\%.

The plots of the probability distribution of the contact network show
that the complete network---in which all possible contacts occur---is
the most likely. However, the probability of this network is lower than
one might expect. We estimate this probability to be 0.49 in households
with two 0--5 year olds and two 19--35 year olds, and 0.36 in
households with two 12--18 year olds and two 36$+$ year olds. The
dependency in contact behavior arising from our model is apparent in
these plots.

We have made some strong assumptions for our model. First, we have
assumed that the only dependence in ties arises from household members
being away from home. Our data suggest that there is indeed an ``away
from home'' effect on contact behavior, but other dependencies are
likely to exist. For example, one parent contacting a child may reduce
the probability that the other parent contacts the child, if one parent
has more child care responsibilities. In addition, our assumption that
the events of members being at home or away from home are independent
is quite strong. Family members are likely to travel together, and in a
household with small children, if one parent is away from home, the
other is probably more likely to be at home. Furthermore, we assumed
that contacts occur independently, conditional on members being at
home. In fact, contact between two family members may influence their
behavior with others, conditional on all of them being at home. We have
also assumed that contact behavior does not change when a household
member is away, other than the removal of contacts to that member. In
fact, it is possible that contact density tends to increase when some
members are away, violating this assumption. Our data do not contain
information to estimate these other potential dependencies. We have
estimated one dependency in contact behavior, informed by the data and
by a~reasonable social theory. Our model is a simplification of the
true underlying process, and further data is required to estimate
additional dependencies and assess whether our model captures the
network structures relevant to the disease transmission process. We
recommend collecting complete network data to analyze these patterns.

Finally, we have assumed that contacts depend on the age categories of
the two members. This assumption is realistic, as evidenced by our
different contact probability estimates in the matrices. However,
contacts could also depend on gender. In particular, mothers may be
more likely to contact children than fathers. Although our data set
contains the gender of each respondent and of all contacts, it does not
contain the gender of each household member. For this reason, including
gender as a predictor is not straightforward.

Our predictions could be improved by collecting additional data. We
recommend asking respondents whether they were at home on the day of
the survey, whether contacted persons were household members, and
whether each household member was at home on the survey day. It could
also be useful to collect the gender of each household member. Based on
our recommendation, the next implementation of POLYMOD in Belgium, as
well as similar studies in Vietnam and Thailand, ask respondents
to identify whether contacted people are household members [Horby et
al. (\citeyear{Horetal})]. In addition, we recommend collection of complete network data
to validate our results and improve understanding of within-household
contact behavior.

Our method can be used in other settings to infer networks from
egocentric data. For example, our method could be used to infer
household contact networks in cultures with larger household sizes than
commonly found in Belgium. A study of household economic networks in a
Malawian village found a mean household size of 9, rather than 3.24 as
in our Belgian data set [Potter and Handcock (\citeyear{PotHan10})]. Our method could
also be used to infer within-classroom networks or within-workplace
networks from the POLYMOD data.

We have demonstrated that this method works reasonably well for small
networks. As the network size increases, the proportion of the network
reported by a single respondent decreases, but identifiability of the
parameter vector depends on the number of age categories. As long as
there is an adequate number of respondents in each age category, the
parameter vector remains identifiable as network size increases.
Computation time is an issue because the number of hidden
configurations increases at a faster rate than network size. The number
of hidden configurations depends on the number of age categories, the
network size, the distribution of household age compositions, the
number of respondents, and the number of reported contacts. Computation
is still feasible for household networks with up to 10 members and for
larger sizes if the number of age categories is reduced. Classroom,
workplace, or daycare networks could be modeled with a single age
category. With a single age category, estimation for networks with up
to 50 members is feasible.

Our method requires a single respondent per network, a common sampling
design for household studies. If multiple respondents per network are
observed, their reports will not be independent, so the joint\vadjust{\eject}
likelihood is not the product of the marginal likelihoods as we
assumed. The independence assumption is reasonable for inference of
small networks when respondents have been sampled at random from an
entire country as in the POLYMOD study. For inference of much larger
networks, with hundreds or thousands of members, it would be more
convenient to sample multiple members per network and develop an
inference technique accounting for the dependence in contact reports.

We have developed a model to infer complete within-household contact
networks from egocentric data. Although our results are from a single
survey, they are broadly relevant to epidemic models. Our model
incorporates dependency in contact behavior by estimating a latent
variable indicating which household members are at home, and our
inferred contact structure departs from the standard random mixing
assumption. In addition, we find higher contact probabilities in
households with 2--3 members than in larger households. This should
also be taken into account when estimating transmission parameters from
household-level data. Finally, many epidemic models assume that all
household members contact each other on a given day, but we find that
the probability of all possible contacts occurring is actually fairly
small. Estimation of contact probabilities and of disease transmission
probabilities is often confounded, since disease outcomes are collected
but detailed information about contact behavior is not. By shedding
light on the contact structure, our work can help disentangle the
contact process from the transmission process. Our findings can be used
to improve epidemic models and estimation methods. As future work, we
propose integrating our findings into these models and performing
simulation studies to evaluate their impact on results.

\begin{appendix}
\section*{Appendix: Proof of identifiability}\label{app}
\setcounter{section}{1}

\begin{thm}[(Identifiability)]
The latent variable model described in Section \ref{subsection:likelihood} is
identifiable.
\end{thm}

\begin{pf}
To see that our model is identifiable, suppose for the sake of
contradiction that two different sets of probability parameters produce
the same probability distribution. Assuming that the two probability
distributions are equal, we will show that the parameterizations must
be identical, which is a contradiction.

We will denote the two different probability parameter vectors
$\mathbf{p}_A$ and~$\mathbf{p}_B$, so the elements of $\mathbf{p}_A$ which are
the at home indicators are denoted $p_{j,A}$ for $j \in\{1,\ldots,5\}
$, and the contact probabilities are denoted $p_{\mathit{rs},A}$ for $r,s \in\{
1,\ldots, 5\}$. The elements of $\mathbf{p}_B$ have analogous notation.
Recall that the observations used to estimate our model parameters
represent households of diverse sizes and age compositions. With an
infinite amount of data, any type of household may be represented in
the data set. Therefore, a household containing only two members, both\vadjust{\eject}
in age category $k$, may be in the data set. Our observed outcome is
the presence or absence of contact to the other member. Keeping our
notation from the description of the likelihood, the observed outcome
is denoted $w_k$ and is equal to either zero or 1. (The other elements
of $w$ are zero since there are no household members in the other age
categories.)
Using our formula for the likelihood and the assumption that
probability distributions are equal under parameterizations $A$ and
$B$, we have
\begin{equation} \label{eq1}
\mathrm{P}(w_k = 1 | \mathbf{p}_A ) = p_{k,A}^2 p_{kk,A} = \mathrm{P}(w_k = 1 | \mathbf{p}_B ) = p_{k,B}^2 p_{kk,B}.
\end{equation}

We want to show that the corresponding elements of $\mathbf{p}_A$ and
$\mathbf{p}_B$ are equal. For this, we will need information from a
different household, one which contains three members in age category
$k$. For this household, the sufficient statistic is again $w_k$, which
can now take on the values 0, 1, or 2. Under our assumptions,
%
\begin{equation} \label{eq2}
\mathrm{P}(w_k = 2 | \mathbf{p}_A) = p_{k,A}^3 p_{kk,A}^2 = \mathrm{P}(w_k = 2 | \mathbf{p}_B) = p_{k,B}^3 p_{kk,B}^2.
\end{equation}

Dividing \eqref{eq2} by \eqref{eq1}, we obtain
%
\begin{equation}\label{eq3}
p_{k,A} p_{kk,A} = p_{k,B} p_{kk,B}.
\end{equation}

Now dividing \eqref{eq1} by \eqref{eq3} yields
%
\begin{equation}\label{eq4}
p_{k,A} = p_{k,B}.
\end{equation}

Thus, we have shown that the ``at home'' probability parameters are the
same under parameterizations $\mathbf{p}_A$ and $\mathbf{p}_B$.
To see that the conditional contact probabilities are also equal,
consider a household containing two members in age categories~$r$ and
$s$, and suppose the respondent is in age category~$r$. Our sufficient
statistic is denoted $w_s$, which can take on values 0 or~1. We have
%
\begin{equation} \label{eq5}
\mathrm{P}(w_s = 1 | \mathbf{p}_A ) = p_{s,A} p_{r,A} p_{\mathit{rs},A} = \mathrm{P}(w_s = 1 | \mathbf{p}_B ) = p_{s,B} p_{r,B} p_{\mathit{rs},B}.
\end{equation}

Since we have already proven that $p_{s,A}= p_{s,B}$ for all age
categories $s$, it follows that $p_{\mathit{rs},A}=p_{\mathit{rs},B}$. Thus, the
parameter vectors $\mathbf{p}_A$ and $\mathbf{p}_B$ are identical.
Since we have contradicted our assumption that they were distinct,
we have proven that our model is identifiable.
\end{pf}
\end{appendix}

\begin{supplement}[id=suppA]
\sname{Supplement A}
\stitle{Contact network parameters estimated separately for the
holiday period versus the nonholiday period, and for 2--3 member
households versus 4$+$ member households}
\slink[doi]{10.1214/11-AOAS474SUPPA}
\slink[url]{http://lib.stat.cmu.edu/aoas/474/Supplement\%20A.pdf}
\sdatatype{.pdf}
\sdescription{We present parameter estimates computed separately for
respondents who reported during the Easter holiday period and during a
nonholiday period. Next we report parameters estimated separately for
households with 2--3 members and those with 4$+$ members.}
\end{supplement}

\begin{supplement}[id=suppB]
\sname{Supplement B}
\stitle{Results from simulation study exploring weak identifiability}
\slink[doi]{10.1214/11-AOAS474SUPPB}
\slink[url]{http://lib.stat.cmu.edu/aoas/474/Supplement\%20B.pdf}
\sdatatype{.pdf}
\sdescription{We present simulation results evaluating weak
identifiability of our parameters in data sets with low
within-household contact rates and low at-home probabilities.}
\end{supplement}

\vspace*{-2pt}
\begin{supplement}[id=suppC]
\sname{Supplement C}
\stitle{R code used for estimation, bootstrapping, and simulation in
``Estimating within-household contact networks from egocentric data''}
\slink[doi]{10.1214/11-AOAS474SUPPC}
\slink[url]{http://lib.stat.cmu.edu/aoas/474/Supplement\%20C.pdf}
\sdatatype{.pdf}
\sdescription{This supplement includes~R code used to perform
estimation, bootstrap confidence intervals, and perform a~simulation
study assessing weak identifiability in households with low contact
rates and low probabilities of being at home.}
\end{supplement}

\vspace*{-2pt}
\section*{Acknowledgments}
We are very grateful to Niel Hens for sharing the Belgian POLYMOD data,
for his careful reading and comments on this paper, and for inviting
two of the authors to the SIMID (Simulation Models of Infectious
Disease Transmission and Control Processes) workshop on infectious
disease modeling and economic evaluation of vaccines. We appreciate the
detailed, thoughtful comments made by three anonymous reviewers as well
as the Associate Editor of this manuscript. Thanks to Nele Goeyvaerts,
James Wood and John Edmunds for providing valuable comments during the
SIMID workshop in Antwerp, 2010. We are also grateful to Martina
Morris, Steven Goodreau, and members of the UW social network modeling
group. We thank the POLYMOD project for providing the data.


%

\printaddresses

\end{document}